\documentclass[12pt, onecolumn, letterpaper]{article}
\usepackage[
	letterpaper, 
	left=0.75in,
	right=0.75in,
	top=0.75in,
	bottom=0.75in
]{geometry}

\usepackage[dvipsnames]{xcolor}

\usepackage[
	colorlinks=true,
	allcolors=black,
	citecolor=gray,
]{hyperref}
\usepackage{caption}
\usepackage{graphicx}
\usepackage{pgf}
\usepackage{tabularx, booktabs, multirow}
\usepackage{amsmath, amsthm, amsfonts, amssymb}

\usepackage{lineno}
\usepackage{setspace}

\usepackage{listings}
\usepackage[affil-it]{authblk}
\usepackage[authoryear, semicolon, sort&compress]{natbib}
\graphicspath{ {./}, {./figures/} }

\newcommand{\fig}[4]{%
	\begin{figure*}[#4]
    \centering\includegraphics[width=#3]{#1}
    \captionsetup{width=\textwidth}\caption{#2}
    \label{fig:#1}
	\end{figure*}
}

\newcommand{\reffig}[2]{\textbf{\hyperref[fig:#1]{Figure \ref*{fig:#1}}#2}}
\newcommand{\refeq}[1]{\textbf{\hyperref[eq:#1]{Equation \ref*{eq:#1}}}}
\newcommand{\reftable}[1]{\textbf{\hyperref[table:#1]{Table \ref*{table:#1}}}}

\title{\textbf{\textsf{ImageMech: From image to particle spring network for mechanical characterization}}}
\author[$\dagger$]{Yuan Chiang}
\author[$\ddagger$,$\mathsection$,$\mathparagraph$]{Ting-Wai Chiu}
\author[$\dagger$,$\|$]{Shu-Wei Chang\thanks{changsw@ntu.edu.tw}}
\affil[$\dagger$]{\small Department of Civil Engineering, National Taiwan University, Taipei, Taiwan}
\affil[$\ddagger$]{Physics Department, National Taiwan University, Taipei, Taiwan}
\affil[$\mathsection$]{Institute of Physics, Academia Sinica, Taipei, Taiwan}
\affil[$\mathparagraph$]{Physics Department, National Taiwan Normal University, Taipei, Taiwan}
\affil[$\|$]{\small Department of Biomedical Engineering, National Taiwan University, Taipei, Taiwan}

\date{\today}

\begin{document}
\maketitle
\sloppy
\onehalfspacing

\section*{Abstract}
{
\sf
The emerging demand for advanced structural and biological materials calls for novel modeling tools that can rapidly yield high-fidelity estimation on materials properties in design cycles. Lattice spring model (LSM), a coarse-grained particle spring network, has gained attention in recent years for predicting the mechanical properties and giving insights into the fracture mechanism with high reproducibility and generalizability. However, to simulate the materials in sufficient detail for guaranteed numerical stability and convergence, most of the time a large number of particles are needed, greatly diminishing the potential for high-throughput computation and therewith data generation for machine learning frameworks. Here, we implement CuLSM, a GPU-accelerated CUDA C++ code realizing parallelism over the spring list instead of the commonly used spatial decomposition, which requires intermittent updates on the particle neighbor list. Along with the image-to-particle conversion tool Img2Particle, our toolkit offers a fast and flexible platform to characterize the elastic and fracture behaviors of materials, expediting the design process between additive manufacturing and computer-aided design. With the growing demand for new lightweight, adaptable, and multi-functional materials and structures, such tailored and optimized modeling platform has profound impacts, enabling faster exploration in design spaces, better quality control for 3D printing by digital twin techniques, and larger data generation pipelines for image-based generative machine learning models.
}

\clearpage
\doublespacing

\section{Introduction}

Materials with complex geometry and multiple constituents can be difficult to predict the mechanical properties, such as elasticity, plasticity, hysteresis, and fracture. The properties are usually coupled with the structure and topology of materials, and in many cases change under different boundary conditions. Classical solid mechanics are highly accurate if the assumptions of homogeneity and small deformation are practical. Problems arise when materials become nonhomogeneous and undergo large deformation. Multiple assumptions and parameter fittings are often required, engendering intensive computational cost and prolonged calibration. In particular, many biomimetic and bioinspired motifs involve complex structures and composite materials by design \citep{wegst2015bioinspired}. For example, birds have hollow and pneumatized bones for avian purpose. Inside the dense and thin exterior, there are hollows with internal reinforcing structures, including ridges, struts, and foams \citep{sullivan2017extreme} (see \reffig{6}{}). These complex structures pose nearly insurmountable challenges to continuum approaches (such as finite element methods, FEMs) since the number of elements required by sufficiently detailed characteristics increases dramatically. To resolve the computational intractability of ultra-high mesh models, homogenization techniques are often adopted to replace the materials at smaller scales by the equivalent larger ones \citep{roters2010overview}. The degrees of freedom therefore decrease correspondingly in favor of the computing capability and desired scalability. However, the process of homogenization inevitably losses information content and geometry details, leading to inaccurate evaluation on the mechanical properties. In this regard, a different perspective is necessary for describing the materials in an efficient approach but without much loss of details.

Lattice spring model (LSM) has been proved to be an effective tool to predict the elasticity, plasticity, and fracture behaviors of metals \citep{buxton2001lattice,chen2014novel}, osteon-inspired cellular composites \citep{libonati2017computational}, and geometrically toughened structural composites \citep{chiang2020geometrically, tsai2021mechanical}. The underlying physics of LSM is simply the truncated potential of springs between particles (or beads), which are the representative volume element (RVE) of the discrete material bodies. The fracture occurs when the length of elongated spring exceeds the critical length, leading to the spring breakage and the release of stored elastic strain energy. This straightforward criteria has provided predictive insights in the fracture behaviors of many brittle materials. 

One of the most popular codes for large-scale particle dynamics simulations is LAMMPS (\href{http://lammps.sandia.gov}{http://lammps.sandia.gov}). The current acceleration of LSM calculation in LAMMPS package relies on spatial decomposition rather than spring list (bond list) parallelization. Parallelization on large spring list is crucial for LSM acceleration since the spring force calculation is the major bottleneck of time integration. Each particle in the middle of triangular packing lattice, for example, has six springs connected with its first nearest neighbors. To enhance the performance of LSM simulation, we develop a CUDA-enhanced lattice spring model code (CuLSM), which implements GPU parallelization on particle and spring lists. CuLSM provides a great speedup for large particle-spring networks with tens of thousands of particles and springs. This work as well as associated codes is important for future large-scale LSM simulations where a large number of particles are necessary to provide enough resolution for biological/biomimetic geometries and complex physical phenomena such as stress concentration, shielding, and plastic zone. 

Here we present a handy platform to evaluate the mechanical properties of biological or biomimetic materials design based on 2D image geometry and prescribed materials constants. The images can be obtained from microscopy, computed tomography scan \citep{bibb2011computed,liang2009macro}, or other imaging methods and artificial design (such as generative adversarial networks) and can be converted into different types of particles based on the gray-scale pixel values. We report an image-particle conversion tool---Img2Particle, which takes the image and number of particle types as input, and outputs the triangular packing particle model with boundary and notch for mechanical characterization. CuLSM subsequently performs displacement-control mechanical test to determine the mechanical properties. System energies, particle trajectories and other derived attributes are computed using various parallelism scheme. The platform provides reliable pipeline from image to mechanical properties and meanwhile achieves high-performance speedup compared with CPU-centered programs.

\fig{6}{ImageMech platform from image to mechanical properties.  Taiwan blue magpie (\textit{Urocissa caerulea}) (photo credit: John\& Fish on Flickr, \href{https://creativecommons.org/licenses/by-nc-nd/2.0/}{CC BY-NC-ND 2.0}). Cross section of bird bone (photo credit: Josef Reischig, \href{https://creativecommons.org/licenses/by-sa/3.0/}{CC BY-SA 3.0}).}{\textwidth}{!htb}

\section{Materials and Methods}

\subsection{Force calculation}

Consider two particles $i,j$ connected by a harmonic spring of stiffness $k$. The potential energy (elastic strain energy) stored in the spring can be expressed in a function of two particle coordinates $\mathbf{r}_i$ and $\mathbf{r}_j$. For a system with $N$ particles and $M$ springs, the total potential energy is \begin{equation}
U(\mathbf{r}) = \sum_{(i, j)\in \mathbf{M}} \frac{1}{2} k \left(r_{ij} - r_{ij}^0\right)^2 \left( 1 - \Xi(r_{ij} - r_c)\right)
\end{equation} where $r_{ij} = \| \mathbf{r}_i - \mathbf{r}_j\|$ and $r_{ij}^0$ are the instantaneous length and equilibrium length of the spring between particles $i,j$. $(i,j)$ is the unique pair of particles in the spring set $\mathbf{M}$ (spring list) of size $M$. $\Xi(r_{ij}-r_c)$ is the Heaviside step function switching on at cutoff $r_c$, where the spring breakage happens.


The force exerted on the individual particle $i$ can be obtained through the gradient of potential energy \begin{equation}
\begin{aligned}
\mathbf{F}^i &= - \frac{\partial U}{\partial \mathbf{r}_i} \\
&= - \sum_{j\in\mathcal{N}(i)} k \left(r_{ij} - r_{ij}^0\right) \frac{\partial r_{ij}}{\partial \mathbf{r}_i} \left( 1 - \Xi(r_{ij} - r_c)\right)\\
&= - \sum_{j\in\mathcal{N}(i)} k \left(r_{ij} - r_{ij}^0\right) 
\frac{ \mathbf{r}_i - \mathbf{r}_j }{\| \mathbf{r}_i - \mathbf{r}_j\|} \left( 1 - \Xi(r_{ij} - r_c)\right) = \sum_{j\in\mathcal{N}(i)} \mathbf{f}^{ji}
\end{aligned}
\end{equation}, where $\mathbf{f}^{ji}$ is the force applied by the spring $(i,j)$ on the particle $i$. 

\subsection{Velocity Verlet integration}

Velocity Verlet integration is used to solve the second-order ODE of Newton's equation of motion $\mathbf{F} = m\ddot{\mathbf{x}}$. One Verlet integration iteration contains three subroutines. First, given positions $\mathbf{x}$, velocities $\mathbf{v}$ as well as accelerations $\mathbf{a}$ of all particles at time $t$, the positions at the next timestep $t+\Delta t$ are calculated as \begin{equation}
\mathbf{x}(t+\Delta t) = \mathbf{x}(t) + \mathbf{v}(t) \Delta t + \frac{1}{2}\mathbf{a}(t)\Delta t^2\end{equation}. Second, the accelerations at the next timestep are obtained from the forces using the configuration at the next timestep $\mathbf{x}(t+\Delta t)$. \begin{equation}
\mathbf{a}(t+\Delta t) = \frac{1}{m}\mathbf{F}\left(\mathbf{x}(t+\Delta t)\right) = -\frac{1}{m}\boldsymbol{\nabla}U\left(\mathbf{x}(t+\Delta t)\right)
\end{equation}. Third, the velocities at the next timestep are then updated as \begin{equation}
\mathbf{v}(t+\Delta t) = \mathbf{v}(t) + \frac{1}{2}\left(\mathbf{a}(t) + \mathbf{a}(t + \Delta t)\right) \Delta t
\end{equation}. In code implementation, we use the half-step velocity scheme to further reduce the memory usage of acceleration vectors. Velocity verlet integration has been proved to be numerically stable and possess important properties for physics such as time reversibility. 

\subsection{GPU parallelization}

Instead of using spatial decomposition which requires \textit{prior} knowledge of particle coordinates and multiple CPU threads to divide entire domain into several computing subdomains, this work applies GPU parallelization to the force calculations of spring list. By doing so, the algorithm focuses on the pair relations between particles connected by springs regardless of their separating distance. The method has a merit that the examination of particle coordinates is unnecessary and therefore accelerates the computing speed.

Simulations are implemented by the in-house CUDA C++ code CuLSM on a desktop with Intel i5-8400 and Nvidia GeForce GTX 1060. First, vectors of positions, velocities, and accelerations of all particles are copied from host to device memory. All of the subsequent boundary displacement and velocity Verlet integration are executed on the device, with periodic callback copying from device to host when the output of particle states are needed. Five GPU kernel functions for boundary displacement, updating position, calculating force, updating acceleration, and updating velocity are implemented at each timestep controlled sequentially by CPU. 

Positions, velocities, and accelerations vectors of all particles are flattened into 1D array and assigned continuously in both host and device memory. 1D block in 1D grid is used, and the block size is fixed as 256 for both particle and spring list. The grid size is dynamically allocated according to the model size of LSM. 

\reffig{1}{} shows the computing flowchart in CuLSM. In the preprocessing stage, the initial particle-spring network is constructed from the desirable geometry. Particle masses and spring parameters are then assigned according to their specific types. After the model is constructed, the boundary conditions and simulation configurations are set. At this stage, CuLSM has read model input, boundary conditions, and simulation configurations and has stored the data in host memory. Before simulation starts, particle and bond vectors are copied from host memory to device memory. At each timestep, boundary displacements are first applied using a GPU kernel function. Another three GPU kernel functions for updating positions, velocities, and accelerations are then initialized for velocity Verlet integration. The position, velocity, and acceleration vectors are flattened into 1D arrays and are allocated continuously in the global memory space. Due to the independence of vector spaces, each thread takes care of single component at a time. However, in the force calculation, the \textit{race condition} emerges when multiple spring forces try to access and add particle forces at the same time, leading to memory conflicts and unexpected results. Thus, the \textit{atomic operation} is used to serializing the requests (access and addition) from threads across the entire grid. The particle forces are first set as zeros and then summed over spring forces using \texttt{atomicAdd} function, as shown in the following code.

\begin{lstlisting}[language=C++]
__global__ void calculate_force(double* x, double* f, float* k, float* r0, float* rc, int* atom_i, int* atom_j, int natoms, int nbonds)
{
    int i = blockIdx.x * blockDim.x + threadIdx.x;
    
    if (i < natoms) {
        f[i*3] = 0;
        f[i*3 + 1] = 0;
        f[i*3 + 2] = 0;
    }
    __syncthreads();

    if (i < nbonds) {

        int ai = atom_i[i];
        int aj = atom_j[i];
        int index_i = ai * 3;
        int index_j = aj * 3;
        
        double r_ij = sqrt(pow(x[index_j] - x[index_i], 2) + 
                          pow(x[index_j + 1] - x[index_i + 1], 2) + 
                          pow(x[index_j + 2] - x[index_i + 2], 2));

        if (r_ij > rc[i]) k[i] = 0;

        double fix = -k[i] * (r_ij - r0[i]) * (x[index_i] - x[index_j])/r_ij;
        double fiy = -k[i] * (r_ij - r0[i]) * (x[index_i + 1] - x[index_j + 1])/r_ij;
        double fiz = -k[i] * (r_ij - r0[i]) * (x[index_i + 2] - x[index_j + 2])/r_ij;

        atomicAdd(&f[index_i], fix);
        atomicAdd(&f[index_i + 1], fiy);
        atomicAdd(&f[index_i + 2], fiz);
        
        atomicAdd(&f[index_j], -fix);
        atomicAdd(&f[index_j + 1], -fiy);
        atomicAdd(&f[index_j + 2], -fiz);
    }

    __syncthreads();
}
\end{lstlisting}

At each iteration, timestep is checked if satisfying the conditions for callback or termination. Once the condition for simulation output is satisfied, particle position and velocity vectors are copied back from device to host memory. The spring stiffness vector is also copied for calculating potential energy. The system potential energy and kinetic energy are calculated on CPU.  

\fig{1}{Computing flowchart in CuLSM. The green blocks are implemented by GPU kernels, which parallelize particle and spring vectors. At each iteration, timestep is checked if satisfying the conditions for callback or termination.}{\textwidth}{!htb}

\clearpage

\section{Results}

\subsection{CuLSM demonstrates strong validity against analytical and numerical results}

We first compare the trajectory of a simple harmonic oscillator solved numerically by CuLSM with the analytical solution. For a system consisting of two particles with mass $m=1$ kg connected by a harmonic spring with spring constant $k=1\times10^{-4}$ N/m and equilibrium distance $r_0=10$m, the equation of motion is a second-order ordinary differential equation: \begin{equation}
m\ddot{x} + kx = kr_0
\end{equation}. We fix one particle at the origin $x=0$m and place another one still at $x=3r_0/2$m when time $t=0$s, as shown in \reffig{2}{A}. The time integral interval $\delta t$ for CuLSM is set as $1$s. The simulation was run for $1000$s and the output interval is $10$s. As depicted by \reffig{2}{B}, our model provides an accurate numerical solution for a simple harmonic oscillator without error accumulation over time.

\fig{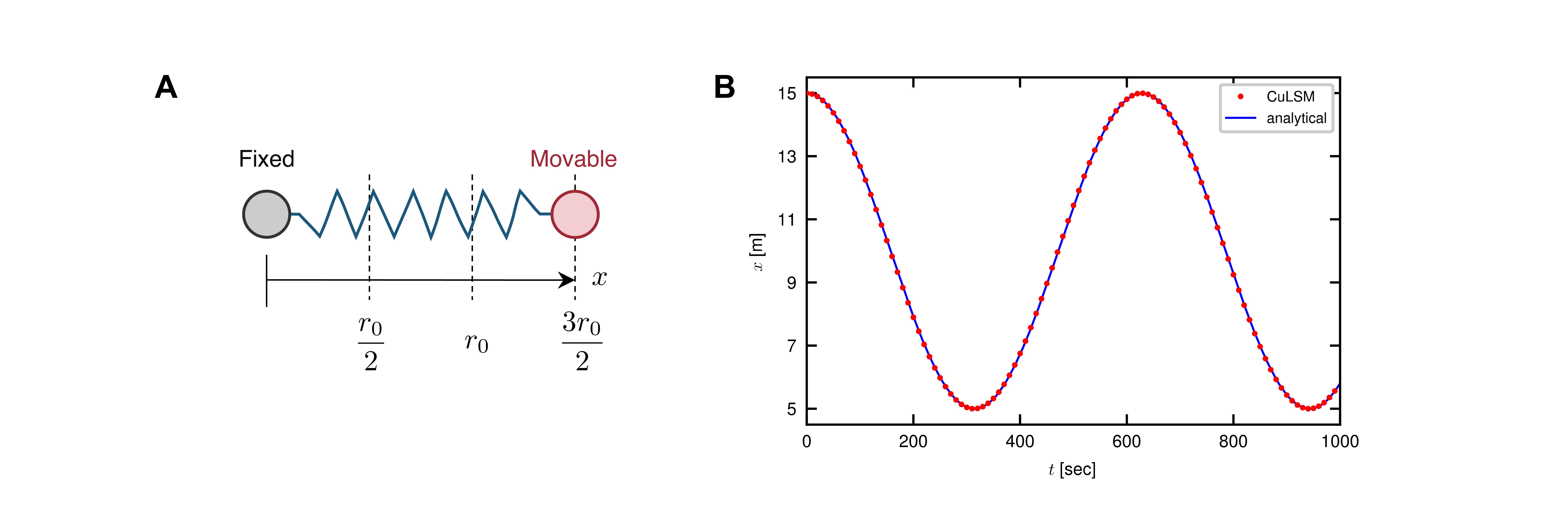}{Validation of CuLSM against the analytical solution of a simple harmonic oscillator. \textbf{(A)} Boundary and initial conditions of the oscillator. \textbf{(B)} Trajectory computed by CuLSM compared to analytical solution.}{\textwidth}{!htb}

\begin{table}[!htb]
\centering
\caption{Constants of linear fracture spring}
\label{table:2}
\begin{tabular}{rccc}
\toprule
\multirow{2}{*}{Springs} & $k_e$ & $r_0$ & $r_c$ \\
& [$\text{M}\text{T}^{-2}$] & [L] & [L] \\
\cmidrule{1-1} \cmidrule(l){2-4} 
stiff-stiff	& $2.00\times10^{-3}$	& 10	& 10.1	\\ 
soft-soft	& $2.00\times10^{-5}$	& 10	& 11.0	\\ 
stiff-soft	& $1.25\times10^{-4}$	& 10	& 10.4	\\ 
\bottomrule
\end{tabular}
\end{table}

We also test our code against LAMMPS (3 Mar 2020, stable release) and compare the performance in the next subsection. As illustrated by \reffig{3}{A}, we construct a series of 2D composite materials with the soft inclusions arranged in a Poisson distribution \citep{chiang2020geometrically}. Three kinds of linear fracture springs, including stiff-stiff, soft-soft, and stiff-soft springs, are used to model stiff, soft and interfacial materials (\reftable{2}). The stiff, soft, and boundary particles are marked as dark blue, light blue, and red, respectively. To model the mode-I fracture behaviors, boundary particles were displaced apart along $x$ axis at the strain rate of $10^{-6}$. The size of composites increases from $1000 \times 1000$ to $2000 \times 2000$ squared unit length, with area ratio $\rho_A$ linearly increasing from $1.0$ to $4.0$ (\reftable{1}). The uniaxial tensile tests are performed to validate the results by CuLSM against those by LAMMPS. As shown in \reffig{3}{B}, the potential and kinetic energies computed by CuLSM perfectly coincide with those computed by LAMMPS before the peaks of potential energies. We also note that the potential and kinetic energies increase as the size of Poisson composite become large. Small energy discrepancies at large strain are observed, but the tendencies are similar. We further compare the fracture patterns obtained from CuLSM and LAMMMPS (\reffig{4}{}). Regardless of the size of the composites, CuLSM and LAMMPS yield akin fracture patterns. The cracks nucleate, propagate, and bifurcate at strikingly similar locations in CuLSM and LAMMPS series, proving strong fidelity of CuLSM. CuLSM reads input of particle geometry from LAMMPS Data file formatted in \textit{bond} atom style. During simulation output, CuLSM outputs particle coordinates in LAMMPS Dump file. The outputted files are readily readable and operable by visualization tools such as OVITO \citep{stukowski2009visualization}.

\fig{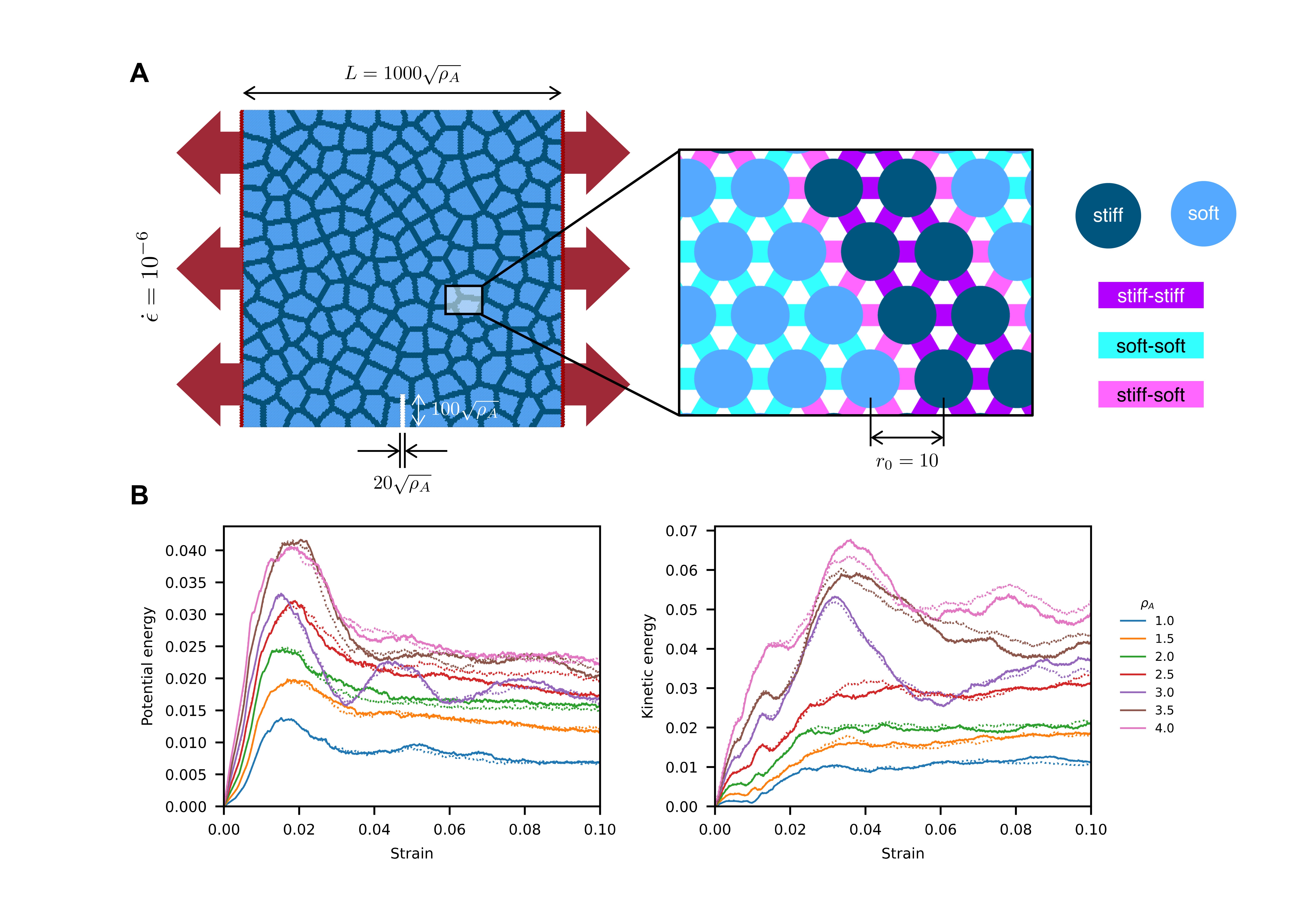}{Mode-I fracture simulation of Poisson composites by LSM. \textbf{(A)} Model size, notch size, and boundary conditions. Stiff-stiff, stiff-soft, and soft-soft springs are used to model stiff, soft, and interfacial materials. \textbf{(B)} Potential and kinetic energies of LSMs with different area ratios $\rho_A$ ranging from $1.0$ to $4.0$. The solid lines are computed by CuLSM, and the dashed lines are computed by LAMMPS.}{\textwidth}{!htb}

\fig{4}{Mode-I fracture patterns predicted by CuLSM and LAMMPS at engineering strain $\epsilon = 0.075$.}{\textwidth}{!htb}


%
%

In \reffig{7}{}, we present virial stress $\boldsymbol{\sigma}^V$ \citep{subramaniyan2008continuum,thompson2009general} and Lagrangian strain $\boldsymbol{\epsilon}^L$ \citep{shimizu2007theory} fields of Poisson composite at bulk engineering strain $\epsilon = 0.02$: \begin{align}
\sigma^V_{ij} &= \frac{1}{\Omega} \sum_{k\in\Omega}\left(\frac{1}{2}\sum_{l\in\Omega}\left(x^{l}_i - x^{k}_i\right)f^{kl}_j - m^k v^k_i v^k_j\right) \\
\epsilon^L_{ij} &= \frac{1}{2}\left(J_{ij}J_{ji} - \delta_{ij}\right)
\end{align}, where $\Omega$ is the finite domain volume considered, $x, v$ are particle position and velocity, $f$ is the force between particle pairs; $J$ is the locally affine transformation matrix considering the relative displacement of particle with its first nearest neighbors, and $\delta$ is the Kronecker delta. The result indicates that the discrepancies of stress and strain fields calculated by CuLSM and LAMMPS are negligible.

\fig{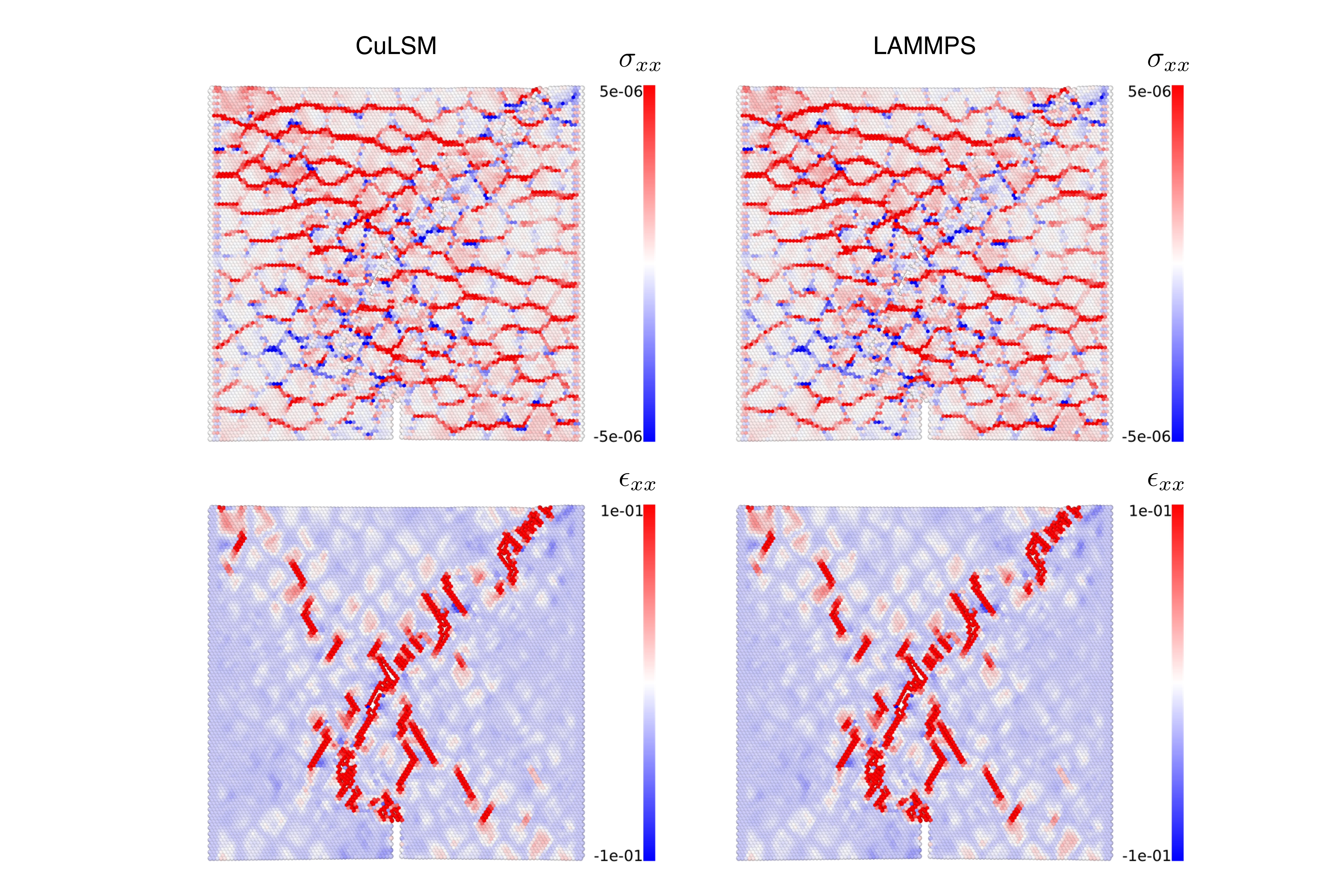}{Stress and strain fields in Poisson composites calculated by CuLSM and LAMMPS at engineering strain $\epsilon = 0.02$.}{\textwidth}{!htb}

\subsection{Benchmarks of CuLSM acceleration}

To benchmark the performance of CuLSM, we record the computing time of mode-I fracture simulations on Poisson composites of different sizes, as listed in \reftable{1}. In \reffig{8}{}, we compare the total wall time of simulations by CuLSM (1 CPU + 1 GPU) and LAMMPS with 1 CPU, 2 CPUs, 4 CPUs, 1 CPU + 1 GPU. With inter-processor communication cutoff $r_\text{comm} = 100 r^0$ and default step interval for neighbor list update $T_n = 10$, LAMMPS with 1 CPU can be one to two orders slower than CuLSM. With these settings, LAMMPS is unfavorably slow and spatial decomposition scheme is incapable of accelerating the LSM simulation efficiently. Note that LAMMPS does not currently support GPU acceleration on bond potentials. Therefore, LAMMPS 1 CPU + 1 GPU shows no speedup compared to LAMMPS 1 CPU. With communication cutoff ($r_\text{comm} = 4 r^0$) and turning off the neighbor  list update ($T_n = \infty$), the total wall time of LAMMPS scales in the same order with CuLSM with respect to particle number. CuLSM can be up to 4.4 times faster than LAMMPS with 1 CPU and around 1.5 speedup compared to LAMMPS with 4 CPUs. CuLSM-CPU with 1 CPU has comparable speed with LAMMPS with 2 CPUs. Note that the optimal neighbor setting depends on the simulation cases for the spatial decomposition scheme. The GPU speedup of CuLSM, \textit{i.e.} the speedup of CuLSM 1 CPU + 1 GPU against CuLSM-CPU 1 CPU, is also presented in the bottom panel of \reffig{8}{}. On the machine with Intel i5-8400 and Nvidia GeForce GTX 1060, the GPU speedup of CuLSM is about 2.5. CuLSM reduces total wall time (including input, output, and copying) by a considerable margin, with only 1 CPU and 1 GPU. The enhanced performance results from the parallelization on particle and spring lists. The input files for all the benchmarks and more information can be found online at the link in Data Availability Statement.

\begin{table}[!htb]
\centering
\caption{Model summary of Poisson composites}
\label{table:1}
\begin{tabularx}{\textwidth}{
	c
	>{\arraybackslash}X
	>{\arraybackslash}X
	>{\arraybackslash}X
} 
\toprule
Area ratio & Size & Particle number & Spring number \\
\cmidrule{1-1} \cmidrule(l){2-4} 
1.0	& $1000\times 1000$	& 12046	& 34753	\\ 
1.5	& $1225\times 1225$	& 17862	& 52104	\\ 
2.0	& $1414\times 1414$	& 23729	& 69476 \\ 
2.5	& $1581\times 1581$	& 29418 & 86080 \\ 
3.0	& $1732\times 1732$	& 35330 & 103904 \\ 
3.5	& $1871\times 1871$	& 41179 & 120960 \\ 
4.0	& $2000\times 2000$	& 46836 & 137759 \\ 
\bottomrule
\end{tabularx}
\end{table}

\fig{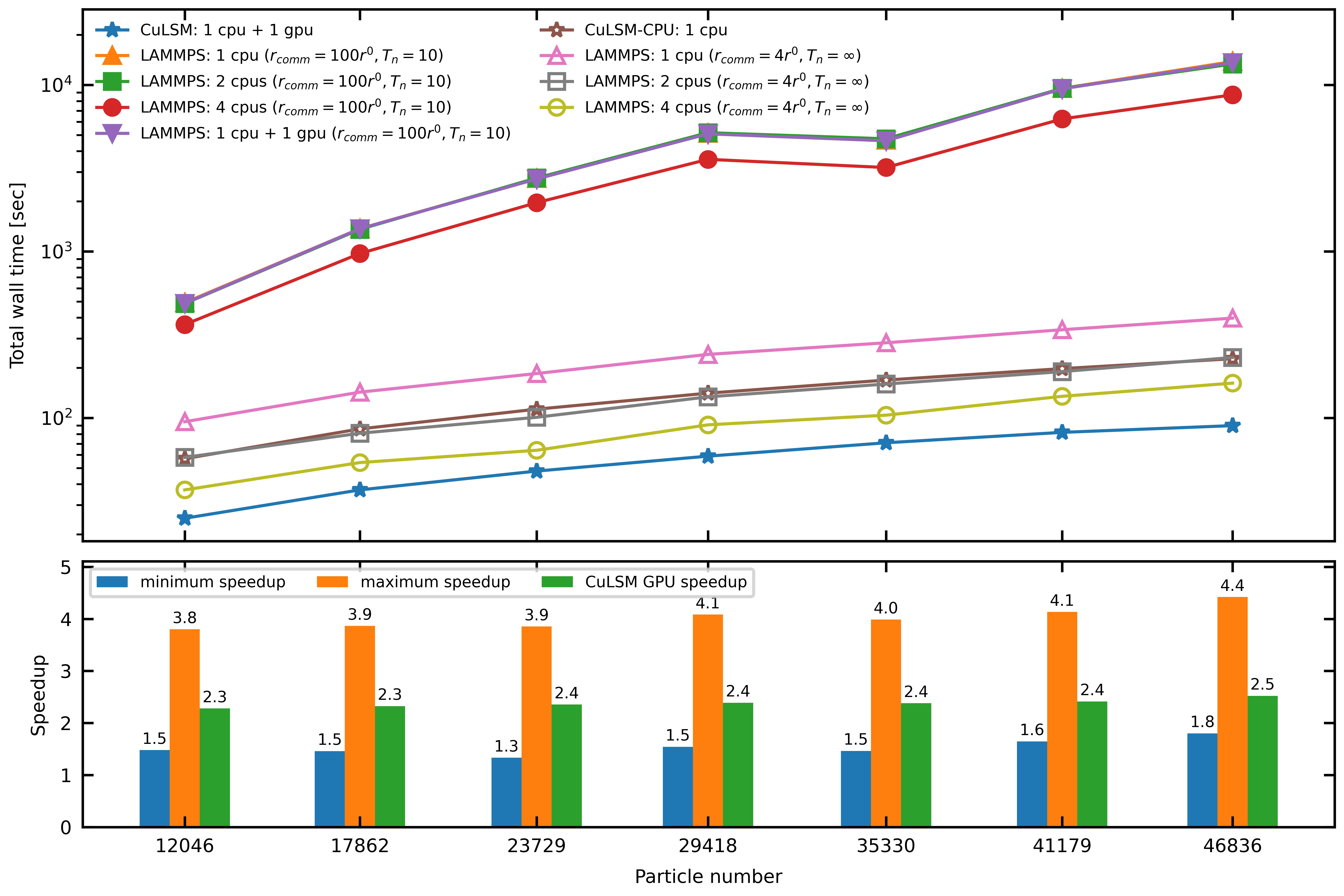}{Comparison of the total computing wall time by CuLSM and LAMMPS. Note that LAMMPS: 1 CPU + 1 GPU does not support parallelism on spring list. The bottom panel compares the maximum and minimum speedup of CuLSM against LAMMPS of $r_\text{comm} = 4 r^0$ and $T_n = \infty$. The maximum speedup compares CuLSM: 1 CPU + 1 GPU with LAMMPS: 1 CPU, and the minimum speedup compares CuLSM: 1 CPU + 1 GPU with LAMMPS: 4 CPUs. The GPU speedup of CuLSM (CuLSM: 1 CPU + 1 GPU versus CuLSM-CPU: 1 CPU) is presented.}{\textwidth}{!htb}

\clearpage
\section{Discussion and Conclusion}

In this work, we present a CUDA C++ code CuLSM for large-scale LSM simulations. By realizing the parallelism on particle and spring lists, CuLSM has been optimized for LSM simulations and secures a remarkable boost in computing speed in comparison with general-purpose LAMMPS package. Since all of the interactions in LSM are harmonic pair potentials, the speedup of spatial decomposition used in LAMMPS is limited. Without updating neighbor list during simulations, CuLSM remarkably accelerates the time integration on GPU and only copy data from device to host when needed. 

Currently, the broken springs are not deleted from the spring list but are irreversibly assigned zero stiffness to emulate the free deformation. We deliberately retain these broken springs since in future studies the stiffness may need to recover when the materials is subject to compression, bending, and cyclic loading. Indeed, different spring properties and mechanical elements, \textit{e.g.}, non-linear elasticity, viscocity, and plasticity, are worth being added into the particle-spring networks. Multi-body potentials such as angle potential and volume-compensated particle method \citep{chen2014novel} are also of interest in future studies. More in-depth theoretical formulations are required for investigating high-level phenomena such as dislocation, Bauschinger effect, and yield surface evolution. 

CuLSM is readily extensible to multi-GPUs and can be further incorporated with multithread environment for the larger and three-dimensional models. To further reduce the memory copying time between host and device, the unified memory can be used to allocate memory address space accessible from any CPUs and GPUs in the system. Moreover, the \textit{parallel reduction} scheme can be adopted to speed up the calculation of the global attributes, such as potential and kinetic energies. 

The future opportunities this work brings include: \begin{itemize}
\item The toolkit we developed provides a faster and more flexible structure-properties platform, which expedites the particle-based simulation and materials design procedures.
\item CuLSM opens the venue for high-throughput and high-fidelity data generation to meet the increasing need for machine learning aided materials design protocols \citep{kim2021deep, sui2021deep}. 
\item The work largely reduces the computational cost for predicting elasticity and fracture behaviors of complex materials systems, further accelerating the design phase through offering predictive insights for additive manufacturing and mechanical experimentation.
\item The LSM simulations provide rich and detailed geometric and topological data where the relationship with high-level mechanical properties underlies. One future research direction would be finding out the physics rules from local to global hierarchy that govern the macroscopic behavior of structural materials. 
\end{itemize}

In conclusion, we provide a powerful and efficient framework to characterize and predict the elasticity and fracture mechanism of materials. The remarkable speedup CuLSM enables entails more extensive application in sophisticated materials design. The robustness and adroitness it promises drive new design perspective other than continuity in practical circumstances. With the emerging new intersection between physics-based simulation and deep learning, the toolkit holds exciting key to advanced materials design.

\bibliography{cudalsm}

\appendix
\section*{Data availability statement}
The CuLSM code associated with this paper is publicly available on GitHub (\href{https://github.com/Chiang-Yuan/culsm}{https://github.com/Chiang-Yuan/culsm}). The Img2Particle code is available on GitHub (\href{https://github.com/Chiang-Yuan/Img2Particle}{https://github.com/Chiang-Yuan/Img2Particle}).

\section*{Author contributions}
YC conceived the idea and developed the research. YC coded the program and performed simulations with advice from TWC. YC wrote the manuscript with input and advice from TWC and SWC. 

\section*{Acknowledgements}
The authors appreciate the financial support from the Ministry of Science and Technology, Taiwan [109-2224-E-007-003, 110-2112-M-003-009, 110-2636-E-002-013]. 

\end{document}